# Graphene enhanced field emission from InP nanocrystals

L. Iemmo[1], A. Di Bartolomeo[1,2,a)], F. Giubileo[2], G. Luongo[1,2], M. Passacantando[3], G. Niu[4], F. Hatami[5], O. Skibitzki[6] and T. Schroeder[6,7]

[1]Physics Department 'E. R. Caianiello', University of Salerno, via Giovanni Paolo II, I-84084, Fisciano, Italy

[2]CNR-SPIN Salerno, via Giovanni Paolo II, I-84084, Fisciano, Italy

[3]Department of Physical and Chemical Science, University of L'Aquila, via Vetoio, I-67100, Coppito, L'Aquila, Italy

[4]Electronic Materials Research Laboratory, Key Laboratory of the Ministry of Education & International Center for Dielectric Research, Xi'an Jiaotong University, Xi'an 710049, People's Republic of China

[5]Institut für Physik, Mathematisch-Naturwissenschaftliche Fakultät, Humboldt Universtät zu Berlin, Newtonstrasse 15, 12489 Berlin, Germany

[6]IHP Microelectronics, Im Technologiepark 25, D-15236 Frankfurt (Oder), Germany

[7]Institute of Physics and Chemistry, BTU Cottbus-Senftenberg, Konrad Zuse Str. 1, D-03046 Cottbus, Germany

[a)]Email: adibartolomeo@unisa.it

**Abstract**

We report the observation of field emission from InP nanocrystals epitaxially grown on an array of p-Si nanotips. We prove that field emission can be enhanced by covering the InP nanocrystals with graphene. The measurements are performed inside a scanning electron microscope chamber with a nano-controlled W-thread used as an anode. We analyze the field emission by Fowler-Nordheim theory and find that the field enhancement factor increases monotonically with the spacing between the anode and the cathode. We also show that InP/p-Si junction has a rectifying behavior, while graphene on InP creates an ohmic contact. Understanding the fundamentals of such nanojunctions is key for applications in nanoelectronics.

Field emission (FE) is a quantum mechanical tunnelling process, which involves extraction of electrons from a conducting solid through a potential barrier, whose width is reduced by the application of an external electric field. Controlled propagation of electrons in vacuum is at the basis of several technological applications, like electron sources,[1] flat panel displays,[2] microwave amplifiers,[3] X-ray sources,[4] etc. For a metallic or semiconducting flat cathode, field emission is enabled by an electric field of the order of several kV/μm. However, if the cathode surface has sharp wedges or protrusions, electrons may be extracted at a considerably lower applied field, because the lines of field converge at the sharp points or edges, and then the physical geometry provides a field enhancement. Several nanostructures have been characterized for their FE properties, like nanodiamonds,[5] carbon nanotubes (CNTs),[6-13] graphene,[14-16] metallic nanowires and nanoparticles of Au,[17] Co, Ni, Cu, Rh,[18] GeSn[19] or semiconducting nanowires of Si,[20] ZnO,[21-22] SiC,[23] GaN,[24] AlN,[25] Ge,[26] etc. To date, no dedicated studies on FE from InP nanocrystals (NCs) have been reported.

Indium phosphide is one of the most popular III-V compound semiconductors with high carrier mobility[27] and direct band gap[28]. Its properties make it of great interest for electronics and optoelectronics. Its availability in the form of NCs, with several edges and corners, can have high potentiality for FE application. Furthermore, since graphene (Gr) is at moment under consideration as promising efficient FE source, it is interesting to study its combined effect with InP-NCs. Graphene has many interesting properties, which can be proven to be useful in FE applications, such as high electron mobility,[29,30] great electric current carrying capacity,[31] high thermal conductivity,[32] record mechanical strength,[33] resilience to high temperatures[34] and humidity,[35] structural flexibility, resistance to molecule diffusion and chemical stability. Due to its high aspect ratio (thickness to lateral size ratio), graphene has high potentiality for FE applications since a dramatically enhanced local electric field is expected at its edges and ripples,[36-41] even though the fabrication of FE devices using graphene edges can be cumbersome. However, being flexible, graphene can be modelled on properly shaped surfaces. In this regards, simply transferring a graphene sheet on an array of epitaxially grown InP-NCs can offer a great geometry configuration for field emission, and simultaneously enhance the resilience of NCs to electrical current and environment chemical interactions.

In this work, we demonstrate that the presence of graphene enhances the FE properties of InP-NCs. We ascribe the improvement to the higher conductivity of graphene as well as to the formation of additional ripples and edges on graphene laid on the InP-NCs array. Furthermore, the presence of the graphene sheet enhances the effective emitting area. We show also that InP/p-Si interface forms a rectifying junction, while graphene establishes an ohmic contact with InP.

We have fabricated and characterized two different devices. The first device consists of an array of InP-NCs, grown by gas-source molecular beam epitaxy on the top of Si-tips patterned on a p-type Si wafer. Fabrication of the Si-tip array includes reactive ion etching of Si, chemical vapour deposition of a $SiO_2$ layer completely covering the Si-tips and chemical mechanical-polishing to reduce the $SiO_2$ thickness till revealing crystalline Si "seeds" of given diameter. The diameters of the base and the opening of Si tips are 500 nm and 50 nm, respectively, and the average size of the InP NCs is ~ 380 nm.[42] The second device was fabricated by transferring commercial CVD graphene from Cu substrates on an InP-NCs/Si-tips array on the same wafer, through a usual wet-transfer process. Further fabrication details can be found in refs.[43-45] In the following, we name as "sample A" the InP-NCs/Si-tips and as "sample B" the Gr/InP-NCs/Si-tips. Scanning electron microscopy (SEM) tilted–view images of the InP-NCs array and Gr/InP-NCs structure are shown in Fig. 1(a) and Fig. 1(b), respectively.

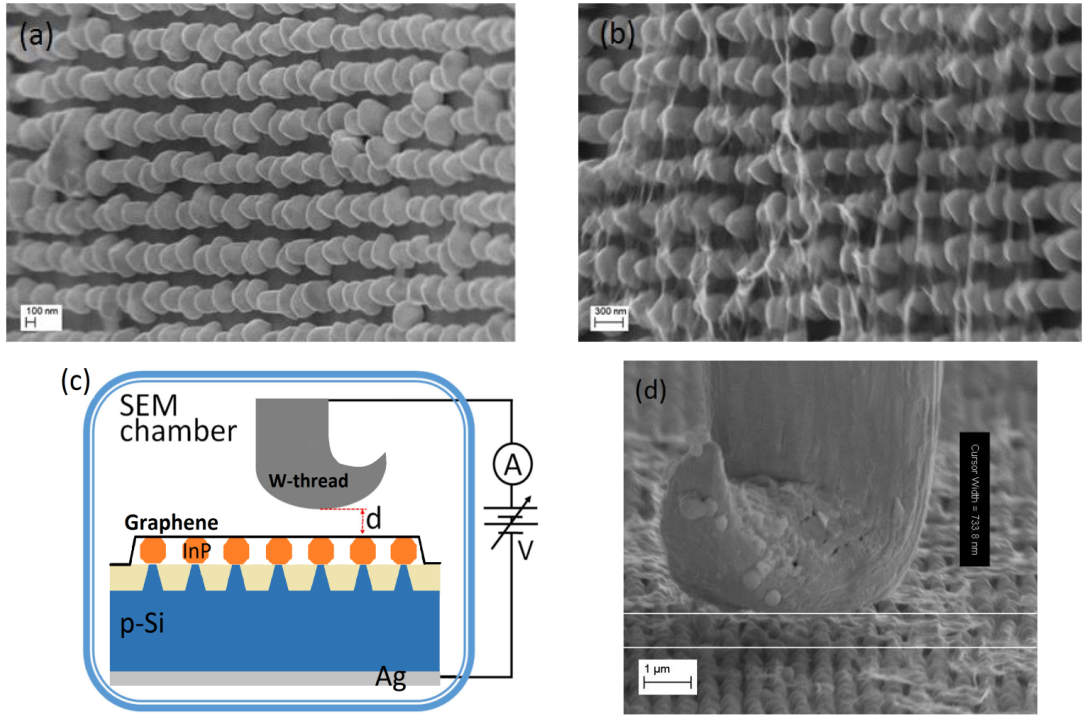

FIG 1. (a) SEM tilted–view image of InP-NCs array (sample A). (b) SEM tilted–view image of Gr/InP-NCs array (sample B). (c) Layout of the Gr/InP-NCs/Si-tips device and field emission measurement setup realized inside a SEM chamber equipped with a W-thread, with fine positioning control over the sample. (d) SEM image of FE device B with the W-thread in the shape of a hook at $d \approx 730$ nm from graphene.

Electrical measurements were performed inside a Zeiss LEO 1430 SEM chamber in high vacuum (< $10^{-6}$ Torr) and at room temperature. A W-thread in the shape of a hook of ~ 4 μm diameter, mounted on a nanoprobe system (manufactured by Kleindeik Company) with two piezoelectric-driven arms, installed inside the SEM chamber, was electrically connected to a semiconductor parameter analyser (Keithley 4200-SCS) working as source-measurement unit (SMU). The circuit configuration for field emission was obtained by retracting the W-thread and adjusting its distance $d$ from the sample surface, while the cathode consisted of a layer of silver painted on the scratched backside of the Si wafer to assure ohmic back-contact. A schematic layout of sample B and the measurement setup is shown in Fig.1(c). The piezoelectric control of the probe thread allows fine tuning of the cathode-anode distance with spatial resolution down to 5 nm. The SEM sample holder was rotated in order to favour the estimation of the thread-sample distance. Voltage sweeps were performed from 0 V to a maximum of 130 V, the voltage limit allowed by the nanomanipulator circuits, and the emitted current was measured with an accuracy better than 0.1 pA. Fig. 1(d) is a SEM image of the W-thread at $d \approx 730$ nm from device B.

In Fig. 2(a) we present current-voltage ($I$-$V$) characteristics obtained with the W-thread in non-physical contact with the sample A and at decreasing distances from it, in the voltage bias range from 0 to 120 V. For $d = 570$ nm, a current starts to flow around 80 V and increases exponentially for about four orders of magnitude over the setup floor noise below $10^{-13}$ A, which is measured at the thread-sample distance of 1 $\mu m$. This behavior is typical of FE current and can be analysed using Fowler-Nordheim (FN) theory,[46] which describes the emitted current with the following equation:

$$I = a \frac{E_S^2}{\Phi} S \cdot exp\left(-b \frac{\Phi^{3/2}}{E_S}\right), \tag{1}$$

where $S$ is the emitting surface area, $E_S$ is the local electric field on that surface, $\Phi$ is the work function of the emitting material and $a$ and $b$ are constants. When $S$ is expressed in cm$^2$, $\Phi$ in eV and $E_S$ in V/cm, the constants $a$ e $b$ assume the value of $1.54 \times 10^{-6}$ AV$^{-2}$eV and $6.83 \times 10^7$ Vcm$^{-1}$eV$^{-3/2}$, respectively. The local electric field at the surface of a nanocrystal can be expressed in terms of the applied potential $V$, the cathode-anode distance $d$ and the field enhancement factor $\beta$ due to its sharp form:

$$E_s = \beta \frac{V}{d}. \tag{2}$$

According to eqs. (1) and (2), $\ln(I/V^2)$ versus $1/V$ is a straight-line (known as the FN plot), whose slope $m = b\Phi^{3/2}d/\beta$ can be used to estimate $\beta$. The inset of Fig. 2(a) shows the corresponding FN plot for $d = 570$ nm. The linearity of the FN plot reveals that the measured current is governed by a conventional FN tunneling. Assuming an InP work function of $\Phi = 4.65$ eV,[47] from the linear fitting we can estimate $\beta \approx 14$.

In Fig. 2(b) we display the evolution of FE current-voltage characteristics for three repeated voltage sweeps obtained with the W-thread at $d = 730$ nm from sample B, in the voltage bias range from 20 to 130 V (for comparison, we add to the FE curve of sample A as well). Despite the higher thread distance (with respect to data reported for sample A in Fig. 2(a)), FE current from sample B starts at remarkably lower applied voltages and attains higher values. Moreover, FE current from sample B raises for seven orders of magnitude in the bias range from 60 V to 130 V and consecutive voltage sweeps show only a slight variation due to electrical conditioning. The initial electrical sweeps on virgin zones have a stabilizing effect and the following I-V characteristics (see Fig. 2(c)) result smoother. As for CNTs,[6] we attribute the electrical conditioning to the desorption of adsorbates (e.g. different types of gases) bond on the graphene surface by weak van der Waals forces. The adsorbates create nanoprotrusions with higher field enhancement factor, which can originate fluctuations in the FE current, and can evaporate at high currents due to the Joule heating, provoking drops and instabilities of the FE current.

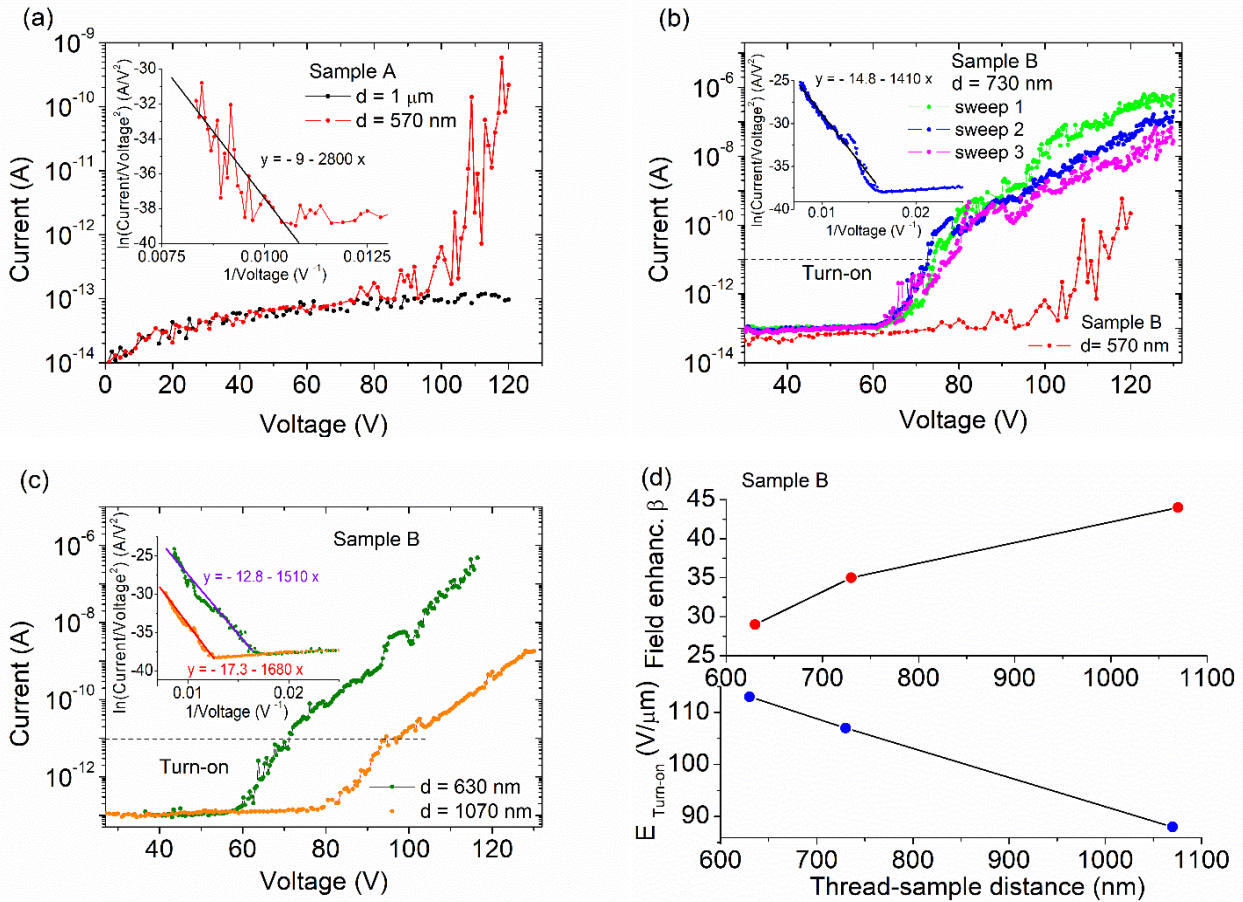

FIG. 2. (a) Current-voltage characteristics measured with the W-thread at given distances from the sample A. At a thread-sample distance of 570 nm, a FE current emerges from the setup floor noise. Inset: FN plot for the sweep at d = 570 nm and its linear fitting (solid line). (b) FE current versus voltage bias characteristics at a distance from the sample A of 570 nm and for three successive sweeps at distance from the sample B of 730 nm. Dashed line identifies the current values considered for the evaluation of the turn-on field. Inset: FN plot for "sweep 2" and its linear fitting. (c) FE curves measured at two different values of the thread-sample B distance. Inset: FN plots for the two sweeps and theirs linear fittings (solid lines). (d) Values of the field enhancement factor and turn-on field versus thread-sample B distance.

The turn-on field, here defined as the applied field necessary to extract a current of 10 pA, is $E_{ON} \approx$ 100 V/μm. We attribute the relatively high turn-on field to the very short thread-sample distance. Indeed, for CNTs it has been shown that the turn-on field increases when the distance decreases.[48] However, the value of the turn-on field is significantly lower or comparable to what has been observed in similar experiments on graphene flakes,[14] carbon nanotubes buckypapers[13] or nanoparticles.[19] In the inset of Fig. 2(b) we show the FN-plot of "sweep 2" for sample B and, from the slope of the fit line, assuming $\Phi = 4.5$ eV as the workfunction of Gr,[49] we extract $\beta \approx 35$. Comparing the field enhancement factor for the samples A and B, estimated at similar distances, we can clearly conclude that graphene FE from InP nanoparticles. An analogous result was found by Zheng et al.[22] on a hybrid graphene-ZnO nanowires device where the addition of a graphene sheet improved the field enhancement factor. Since the parameter $\beta$ depends on the aspect ratio and the spatial distribution of the emitters, they concluded that the pyramidal apices formed by graphene on ZnO nanowires can be sharper than those of ZnO tips, thus causing enhanced FE. In our case, as the graphene sheet on the InP-NCs/Si-tips array has ripples and voids (see Fig. 1(b)), we believe that FE is enhanced also by graphene wrinkles and edges other than by the formation of sharper graphene apices on the InP-NCs. Fig. 2(c) shows the FE $I$-$V$ curves obtained for two different values of the thread-sample B distance (630 nm and 1070 nm). As expected, for reduced distance, $d = 630$ nm, the FE starts at lower applied voltage (~ 55 V compared to ~ 75 V at $d = 1070$ nm). Fig. 2(d) summarizes the values of the field enhancement factor and turn-on field as a function of the thread-sample distance. We found that $E_{ON}$ decreases while $\beta$ increases monotonically with the spacing between the anode and the cathode, similarly to what has been reported for CNT emitters.[6,48,50]

In Fig. 3(a), we report $I$-$V$ characteristics measured with the W-thread in physical contact with both samples A and B. Black sweep refers to the W/Si junction which is obtained by pushing the W-thread on sample A, till letting it slip on surface. SEM imaging shows that the sliding thread, under pressure, easily removes InP-NCs, thus establishing an intimate contact with the Si-tips. The corresponding $I$-$V$ curve has a symmetric behaviour, indicating an ohmic W/Si junction (black curve). The high work function of tungsten or tungsten oxide (the air exposed W-thread can be slightly oxidized), with measured values greater than 5 eV,[51,52] makes the W Fermi level aligned below the valence band of Si, thus resulting in absence of Schottky barrier.[53-55] The same consideration applies to W/InP contact, for which we expect an ohmic behaviour as well. Instead, the other $I$-$V$ curves in Fig. 3(a) exhibit a clear rectifying behavior, implying that the W/InP/Si (red curve) and W/Gr/InP/Si (green curve) heterostructures include Schottky junctions. The forward current at negative voltages suggests that the Schottky junction is formed with the p-type Si substrate. Then it is likely that the rectification is due to the InP/Si interface in the W/In-P/p-Si structure. Otherwise, the similar behaviour of W/Gr/In-P/p-Si system strongly indicates that graphene forms an ohmic contact on InP. Figs. 3(b) and (c) schematize the energy band diagrams at equilibrium of the W/p-Si junction and of the Gr/InP/p-Si heterostructure. Furthermore, from Fig. 3(a), we observe that the presence of graphene increases the current. This is due to the enveloping of the NCs by graphene that increases the contact area thus reducing the overall series electrical resistance. These results are consistent with previous studies on optoelectronic devices, fabricated with the same InP/Si-array process, measuring a Schottky barrier height of 0.13 eV for the InP/p-Si junction.[43]

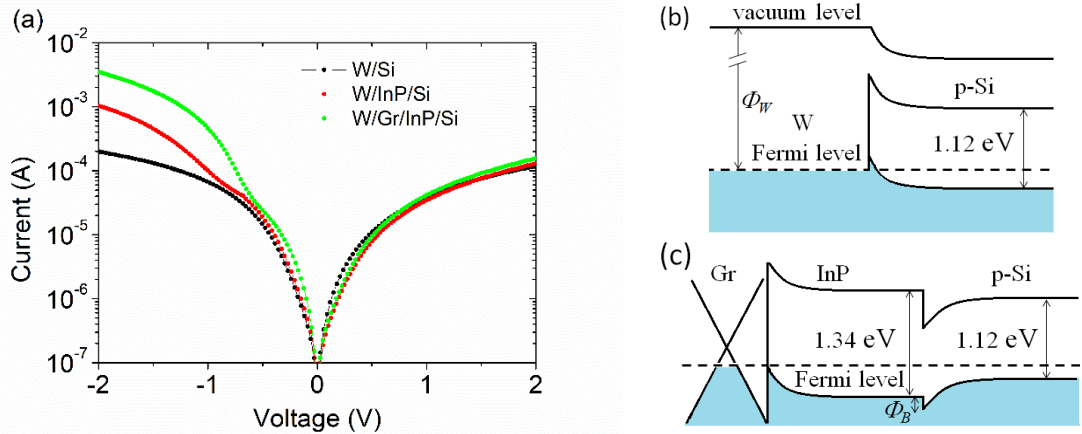

FIG. 3. (a) I-V characteristics measured with the W-thread in physical contact with Si (black curve), InP/Si heterostructure (red curve) and Gr/InP/Si heterostructure (green curve). (b) Energy band diagram at equilibrium of the W/p-Si junction. $\Phi_W > 5$ eV is the work function of W/WO$_x$. (c) Energy band diagram at equilibrium of the Gr/InP/p-Si heterostructure.

In conclusion, we have reported the observation of FE current from InP-NCs epitaxially grown on nanometer–sized Si-tips and proved that such current can be enhanced by graphene on the InP/p-Si heterostructure. We have shown that FE current follows the FN model over seven orders of magnitude. We also gave an estimation of relevant parameters as field enhancement factor and turn-on field. A great advantage of our system with respect to nanowires or nanotube based applications is the regular and accurate positioning of InP nanocrystals, which are selectively grown by epitaxy on the Si-tip array. We have also shown that the InP/p-Si junction forms a rectifying junction and graphene on InP-NCs creates an ohmic contact. The FE capabilities of the hybrid Gr/InP-NCs/Si structure highlighted in this paper can pave the way for new promising applications of InP-NCs in vacuum nano-electronics.

We acknowledge the economic support of POR Campania FSE 2014-2020, Asse III Ob.Specifico l4, Avviso pubblico decreto dirigenziale n. 80 del 31/05/2016, and L.R. num. 5/2002 Finanziamento progetti annualità 2008, Prot. 2014, 0293185, 24/04/2014, and Natural Science Fundamental research

[1] N. de Jonge, M. Allioux, J.T. Oostveen, K.B.K. Teo, and W.I. Milne, Physical Review Letters **94**, (2005).

[2] Q.H. Wang, M. Yan, and R.P.H. Chang, Applied Physics Letters **78**, 1294 (2001).

[3] G. Fursey, *Field Emission in Vacuum Microelectronics* (Kluwer Academic/Plenum Publishers, New York, 2005).

[4] Y. Saito, K. Hata, A. Takakura, J. Yotani, and S. Uemura, Physica B: Condensed Matter **323**, 30 (2002).

[5] M.L. Terranova, S. Orlanducci, M. Rossi, and E. Tamburri, Nanoscale **7**, 5094 (2015).

[6] A. Di Bartolomeo, A. Scarfato, F. Giubileo, F. Bobba, M. Biasiucci, A.M. Cucolo, S. Santucci, and M. Passacantando, Carbon **45**, 2957 (2007).

[7] R.C. Smith, D.C. Cox, and S.R.P. Silva, Applied Physics Letters **87**, 103112 (2005).

[8] M. Passacantando, F. Bussolotti, S. Santucci, A. Di Bartolomeo, F. Giubileo, L. Iemmo, and A.M. Cucolo, Nanotechnology **19**, 395701 (2008).

[9] F. Giubileo, A.D. Bartolomeo, A. Scarfato, L. Iemmo, F. Bobba, M. Passacantando, S. Santucci, and A.M. Cucolo, Carbon **47**, 1074 (2009).

[10] Y. Saito and S. Uemura, Carbon **38**, 169 (2000).

[11] F. Giubileo, A. Di Bartolomeo, M. Sarno, C. Altavilla, S. Santandrea, P. Ciambelli, and A.M. Cucolo, Carbon **50**, 163 (2012).

[12] J.-M. Bonard, K.A. Dean, B.F. Coll, and C. Klinke, Physical Review Letters **89**, 197602 (2002).

[13] F. Giubileo, L. Iemmo, G. Luongo, N. Martucciello, M. Raimondo, L. Guadagno, M. Passacantando, K. Lafdi, and A. Di Bartolomeo, Journal of Materials Science **52**, 6459 (2017).

[14] S. Santandrea, F. Giubileo, V. Grossi, S. Santucci, M. Passacantando, T. Schroeder, G. Lupina, and A. Di Bartolomeo, Applied Physics Letters **98**, 163109 (2011).

[15] S. Kumar, G.S. Duesberg, R. Pratap, and S. Raghavan, Applied Physics Letters **105**, 103107 (2014).

[16] A. Di Bartolomeo, F. Giubileo, L. Iemmo, F. Romeo, S. Russo, S. Unal, M. Passacantando, V. Grossi, and A.M. Cucolo, Applied Physics Letters **109**, 023510 (2016).

[17] Z. Geng-Min, E. Roy, L. Hong-Wen, L. Wei-Min, H. Shi-Min, K. Yu-Zhang, and X. Zeng-Quan, Chinese Physics Letters **19**, 1016 (2002).

[18] S. Xavier, S. Mátéfi-Tempfli, E. Ferain, S. Purcell, S. Enouz-Védrenne, L. Gangloff, E. Minoux, L. Hudanski, P. Vincent, J.-P. Schnell, D. Pribat, L. Piraux, and P. Legagneux, Nanotechnology **19**,

215601 (2008).

[19] A. Di Bartolomeo, M. Passacantando, G. Niu, V. Schlykow, G. Lupina, F. Giubileo, and T. Schroeder, Nanotechnology **27**, 485707 (2016).

[20] M. Choueib, R. Martel, C.S. Cojocaru, A. Ayari, P. Vincent, and S.T. Purcell, ACS Nano **6**, 7463 (2012).

[21] S.-Y. Kuo and H.-I. Lin, Nanoscale Research Letters **9**, 70 (2014).

[22] W.T. Zheng, Y.M. Ho, H.W. Tian, M. Wen, J.L. Qi, and Y.A. Li, The Journal of Physical Chemistry C **113**, 9164 (2009).

[23] K. Senthil and K. Yong, Materials Chemistry and Physics **112**, 88 (2008).

[24] E. Li, G. Wu, Z. Cui, D. Ma, W. Shi, and X. Wang, Nanotechnology **27**, 265707 (2016).

[25] F. Chen, X. Ji, and Q. Zhang, Journal of Alloys and Compounds **646**, 879 (2015).

[26] L. Li, X. Fang, H.G. Chew, F. Zheng, T.H. Liew, X. Xu, Y. Zhang, S. Pan, G. Li, and L. Zhang, Advanced Functional Materials **18**, 1080 (2008).

[27] A.H. Moore, M.D. Scott, J.I. Davies, D.C. Bradley, M.M. Faktor, and H. Chudzynska, Journal of Crystal Growth **77**, 19 (1986).

[28] I. Vurgaftman, J.R. Meyer, and L.R. Ram-Mohan, Journal of Applied Physics **89**, 5815 (2001).

[29] K.S. Novoselov, Science **306**, 666 (2004).

[30] X. Du, I. Skachko, A. Barker, and E.Y. Andrei, Nature Nanotechnology **3**, 491 (2008).

[31] J. Moser, A. Barreiro, and A. Bachtold, Applied Physics Letters **91**, 163513 (2007).

[32] A.A. Balandin, S. Ghosh, W. Bao, I. Calizo, D. Teweldebrhan, F. Miao, and C.N. Lau, Nano Letters **8**, 902 (2008).

[33] C. Lee, X. Wei, J.W. Kysar, and J. Hone, Science **321**, 385 (2008).

[34] J.H. Los, K.V. Zakharchenko, M.I. Katsnelson, and A. Fasolino, Physical Review B **91**, 045415 (2015).

[35] P.-G. Su and C.-F. Chiou, Sensors and Actuators B: Chemical **200**, 9 (2014).

[36] S.W. Lee, S.S. Lee, and E.-H. Yang, Nanoscale Research Letters **4**, 1218 (2009).

[37] A. Malesevic, R. Kemps, A. Vanhulsel, M.P. Chowdhury, A. Volodin, and C. Van Haesendonck, Journal of Applied Physics **104**, 084301 (2008).

[38] M. Qian, T. Feng, H. Ding, L. Lin, H. Li, Y. Chen, and Z. Sun, Nanotechnology **20**, 425702 (2009).

[39] J. Liu, B. Zeng, Z. Wu, J. Zhu, and X. Liu, Applied Physics Letters **97**, 033109 (2010).

[40] Z. Lu, W. Wang, X. Ma, N. Yao, L. Zhang, and B. Zhang, Journal of Nanomaterials **2010**, 1 (2010).

[41] Z.-S. Wu, S. Pei, W. Ren, D. Tang, L. Gao, B. Liu, F. Li, C. Liu, and H.-M. Cheng, Advanced

Materials **21**, 1756 (2009).

[42] G. Niu, G. Capellini, M. A. Schubert , T. Niermann, P. Zaumseil, J. Katzer, H.-M. Krause, O. Skibitzki, M. Lehmann, Y.-H. Xie, H. von Känel and T. Schroeder, Scientific Reports **6**, 22709 (2016).

[43] G. Niu, G. Capellini, F. Hatami, A. Di Bartolomeo, T. Niermann, E.H. Hussein, M.A. Schubert, H.-M. Krause, P. Zaumseil, O. Skibitzki, G. Lupina, W.T. Masselink, M. Lehmann, Y.-H. Xie, and T. Schroeder, ACS Applied Materials & Interfaces **8**, 26948 (2016).

[44] O. Skibitzki, I. Prieto, R. Kozak, G. Capellini, P. Zaumseil, Y. A. Rojas Dasilva, M. D. Rossell, R. Erni, H. von Känel, and T. Schroeder, Nanotechnology **28**, 135301 (2017).

[45] G. Niu, G. Capellini, M.A. Schubert, T. Niermann, P. Zaumseil, J. Katzer, H.-M. Krause, O. Skibitzki, M. Lehmann, Y.-H. Xie, H. von Känel, and T. Schroeder, Scientific Reports **6**, (2016).

[46] R.H. Fowler and L. Nordheim, Proceedings of the Royal Society A: Mathematical, Physical and Engineering Sciences **119**, 173 (1928).

[47] T.E. Fischer, Physical Review **142**, 519 (1966).

[48] R.C. Smith and S.R.P. Silva, Journal of Applied Physics **106**, 014314 (2009).

[49] G. Giovannetti, P.A. Khomyakov, G. Brocks, V.M. Karpan, J. van den Brink, and P.J. Kelly, Physical Review Letters **101**, 026803 (2008).

[50] C.J. Edgcombe and U. Valdre, Journal of Microscopy **203**, 188 (2001).

[51] E.W. Müller, Journal of Applied Physics **26**, 732 (1955).

[52] M. Vasilopoulou, A. Soultati, P. Argitis, T. Stergiopoulos, and D. Davazoglou, The Journal of Physical Chemistry Letters **5**, 1871 (2014).

[53] A. Di Bartolomeo, Physics Reports **606**, 1 (2016).

[54] F. Giubileo and A. Di Bartolomeo, Progress in Surface Science **92**, 143 (2017).

[55] A. Di Bartolomeo, F. Giubileo, G. Luongo, L. Iemmo, N. Martucciello, G. Niu, M. Fraschke, O. Skibitzki, T. Schroeder, and G. Lupina, 2D Materials **4**, 015024 (2016).